\documentclass[aps, prb, reprint, amsmath,amssymb]{revtex4-1}

\usepackage{graphicx}
\usepackage{dcolumn}
\usepackage{bm}
\usepackage{subfig}
\usepackage{caption}
\begin{document}

\title{Landau-Zener population control and dipole measurement of a two level system bath}


\author{M. S. Khalil}
\affiliation{ 
Laboratory for Physical Sciences, College Park, MD, 20740
}
\affiliation{ 
Department of Physics, University of Maryland, College Park, MD, 20742
}

\author{S. Gladchenko}%
\affiliation{ 
Laboratory for Physical Sciences, College Park, MD, 20740
}
\affiliation{ 
Department of Physics, University of Maryland, College Park, MD, 20742
}
\author{M. J. A. Stoutimore}%
\affiliation{ 
Laboratory for Physical Sciences, College Park, MD, 20740
}
\affiliation{ 
Department of Physics, University of Maryland, College Park, MD, 20742
}

\author{F. C. Wellstood}
\affiliation{ 
Department of Physics, University of Maryland, College Park, MD, 20742
}
\affiliation{ 
Joint Quantum Institute, University of Maryland, College Park, MD, 20742
}

\author{A. L. Burin}%
\affiliation{ 
Department of Chemistry, Tulane University, New Orleans, LA, 70118
}

\author{K. D. Osborn}%
 \email{osborn@lps.umd.edu}
\affiliation{ 
Laboratory for Physical Sciences, College Park, MD, 20740
}
\date{\today}

\begin{abstract}
Tunneling two level systems (TLS), present in dielectrics at low temperatures, have been recently studied for fundamental understanding and superconducting device development. According to a recent theory by Burin \textit{et al.}, the TLS bath of any amorphous dielectric experiences a distribution of Landau-Zener transitions if exposed to simultaneous fields. In this experiment we measure amorphous insulating films at millikelvin temperatures with a microwave field and a swept electric field bias using a superconducting resonator. We find that the maximum dielectric loss per microwave photon with the simultaneous fields is approximately the same as that in the equilibrium state, in agreement with the generic material theory. In addition, we find that the loss depends on the fields in a way which allows for the separate extraction of the TLS bath dipole moment and density of states. This method allows for the study of the TLS dipole moment in a diverse set of disordered films, and provides a technique for continuously inverting their population.
\end{abstract}

\maketitle 

In quantum computing, two-level systems (TLS) in dielectrics have been found to function as an environmental bath for superconducting quantum elements \cite{Martinis2005,Simmonds2004,Constantin2009,Gao2008} and as quantum memory bits in a hybrid quantum computer \cite{Neeley2008}. The environmental impact of the deleterious bath has led to improved materials \cite{Oh2006,Cicak2010,Paik2010} for superconducting qubits. In recent qubit designs\cite{Geerlings2012,Paik2011}  the geometrical architecture allows only for a small amount of electrical energy storage in the deleterious amorphous metal oxides. Over four decades ago, a now standard model of TLS was introduced which describes charged nanoscale systems moving independently in a distribution of double well potentials, presumably created by undercoordinated bonds \cite{Phillips1972,Anderson1972}. Recent measurements of individual TLSs under application of a strain field are in agreement with this model \cite{Grabovskij2012}. Although the TLS effects are generally known, the precise identity of the atomic defects or bonds that comprise the TLS and dipole moments from a given material are generally not known \cite{Pohl2002,Southworth2009,Queen2013}. Furthermore, it was found that the sudden application of strain or electric fields can result in an immediate change in the TLS density, followed by a slow glassy relaxation to the equilibrium state \cite{Salvino1994,Rogge1996,Natelson1998}, possibly caused by weak TLS-TLS interactions \cite{Carruzzo1994,Burin1995}.

In the case of resonant microwave measurements, the loss tangent is proportional to the weighted TLS density--the TLS density times the dipole moment squared. Experiments on individual TLS provide important quantum properties \cite{Palomaki2010,Neeley2008,Shalibo2010,Stoutimore2012}, but have previously been restricted to an alumina tunneling barrier and must characterize many TLS, one at a time, in order to extract an average dipole moment of the film. The Landau-Zener effect has been used to study a wide variety of qubit systems, including superconducting circuits \cite{Oliver2005,Berns2008,Sillanpaa2006}, silicon-dopants \cite{Dupont-Ferrier2013}, and quantum dots \cite{Cao2013}. A recent theory using this effect predicts that TLS can be characterized using the quantum dynamics created by two simultaneous fields \cite{Burin2013}. Experimental realization of this theory, discussed below, reveals new information about the dynamics of the tunneling systems in amorphous films.

Here we describe measurements of the high-frequency $(\hbar\omega \gg k_BT)$ loss tangent of amorphous PECVD deposited $\mathrm{Si_3N_4}$ films \cite{Paik2010} in a non-equilibrium regime. This regime is reached by sweeping an electric field bias while probing the loss with a microwave (ac) field at millikelvin temperatures. As expected for TLS-laden films, the loss tangent decreases as the microwave power increases. However, we also found that with a sufficiently large bias sweep rate, the loss tangent in the non-equilibrium regime recovers the value of its linear-response steady-state measurement. We compared our loss tangent measurements at high sweep rates to the newly-proposed model, based on Landau-Zener dynamics of a conventional TLS distribution \cite{Burin2013}. By confirming this theory experimentally, we show that the standard TLS model is appropriate for studying a new non-equilibrium regime in amorphous solids, and also show a new method for extracting the TLS dipole moment. Agreement with the theory implies that the population of the TLS bath can be controlled.

\begin{figure*}
\includegraphics[trim = 0cm 0cm 0cm 0cm, clip, scale=0.6]{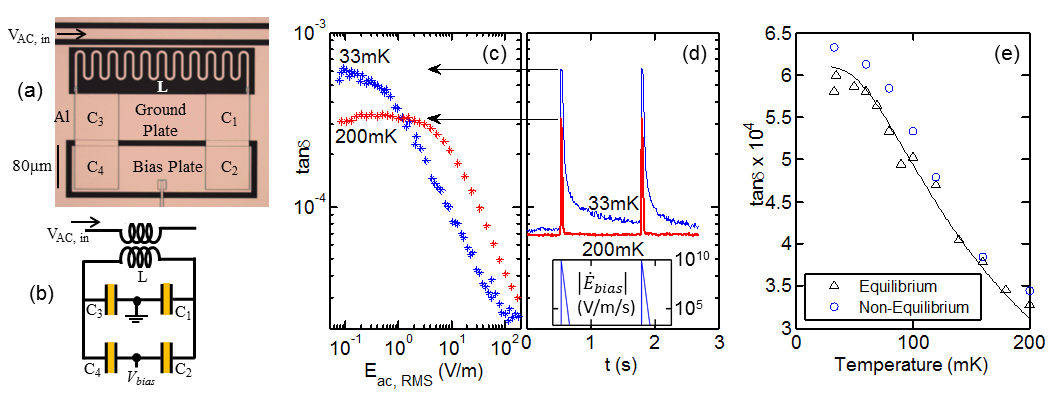}
\caption{\label{fig:TempDep} (color online) (a) Image of biased bridge resonator used. (b) Schematic of the resonator. (c) Steady-state equilibrium loss tangent, measured as a function of the microwave field at temperatures of 33 and 200 mK. (d) Time-dependent non-equilibrium loss tangent measured (main panel) with the electric field sweep rate magnitude (inset). (e) Comparison of linear-response equilibrium loss tangent (black triangles) to the maximum non-equilibrium loss tangent (blue circles) as a function of temperature. The solid curve shows a fit of the conventional theory of thermally saturated TLS, $\tan\delta_0\tanh(\hbar\omega/2k_BT)$, to the equilibrium data.}
\end{figure*}

Measurements were made with a thin-film superconducting aluminum 4.7 GHz resonator (see Fig. \ref{fig:TempDep}(a)) composed of a meandering inductor and four 250 nm thick amorphous $\mathrm{Si_3N_4}$-dielectric parallel-plate capacitors in an electrical bridge design (see Fig. \ref{fig:TempDep}(b)). Arms of the bridge are nominally identical, and a lead allows application of a voltage bias $V_{bias}$ creating a DC voltage difference of $V_{bias}/2$ across each capacitor. We apply resonant microwaves to the system via a coplanar waveguide transmission line that couples the microwave fields into the capacitors. The resonator coupled to the transmission line creates a notch filter, but it is nominally uncoupled from the bias line at resonance due to the balanced electrical design. From measurements of the microwave transmission through the coplanar waveguide \cite{Khalil2012}, we extract the internal quality factor $Q_i=1/\tan\delta$, equal to the inverse  loss tangent of the dielectric films, and the coupling quality factor of 6500. The device is mounted in a sealed copper box attached to the mixing chamber of a dilution refrigerator and measured at 33-200 mK. Filtered transmission lines and a cold low-noise HEMT amplifier allow resonator measurements with less than a single average photon excitation.

The standard TLS model assumes a broad number density distribution, $d^{2}n=d\Delta d\Delta_0 P_0/\Delta_0$, of double well TLS with energy $E_{TLS} = \sqrt{\Delta^2+\Delta_0^2}$, dependent on the tunneling energy $\Delta_0$ and asymmetry energy $\Delta$. The model yields a dielectric loss tangent $\tan\delta=\tan\delta_0 \tanh\left(\hbar\omega/2k_BT\right)/\sqrt{1+\left(E_{ac}/E_c\right)^2}$ \cite{VONSCHICKFUS1977}, where $k_B$ is the Boltzmann constant, $T$ is the temperature, $\tan\delta_0=(\pi P_0p^2)/(3\epsilon)$, $P_0$ is the TLS spectral and spatial density, $p$ is the dipole moment of the TLS, $\epsilon$ is the dielectric permittivity, $E_{ac}$ is the ac field amplitude, and $E_c=\hbar/(p\tau)$. $\tau$ is a characteristic TLS lifetime that depends on the decoherence limit, such that $\tau=\sqrt{T_{1,min}T_2/3}$ for constant coherence time $T_2$ and $\tau=8\sqrt{T_{1,min}T_{2,min}}/(3\pi)$ for the spontaneous emission limit, $T_2=2T_1$; $T_1=(E_{TLS}/\Delta_0)^2T_{1,min}$ is the relaxation time. The loss tangent is mainly sensitive to TLS resonant dynamics near $\hbar\omega$. At low field amplitudes the linear equilibrium response $\tan\delta=\tan\delta_0$ is determined from Fermi's Golden Rule, while at moderate fields the maximum Rabi frequency, $\Omega_{R0}=pE_{ac}/\hbar$, exceeds the TLS decoherence rate, $\Omega_{R0}>>1/\tau$, such that saturation occurs. Figure \ref{fig:TempDep}(c) shows $\tan\delta$ at zero field bias at two temperatures as a function of the RMS field $E_{ac, RMS}=E_{ac}/\sqrt{2}$. As expected, it decreases as the $E_{ac}$ increases and the low ac-field $\tan\delta$ decreases as $T$ increases. The $T$ dependence of the $\tan\delta$ is caused both by thermal saturation of the TLS and by changes in $\tau$ dependent on $T$.

According to the standard double-well model, the bias field $E_{bias}$ should adjust the asymmetry energy between the wells ($\Delta\rightarrow\Delta+2\vec{p}\cdot\vec{E}_{bias}$). Application of fixed voltage biases (data shown in Fig. \ref{fig:TempDep}(c) were taken with zero bias voltage) produce detectable but small variations of order 1\% in the equilibrium loss tangent. These small changes are expected in the standard model distribution because the change in asymmetry results in the same TLS population near resonance within statistical variations.

In contrast, when we drive the resonator with sufficient microwave amplitude to saturate TLS in steady state we find that the loss tangent is sensitive to sufficiently rapid changes in the bias voltage. To observe how the sweep rate affects the loss, we measure the resonator response as a function of time while applying a square waveform, low-pass filtered with a single time constant of 8.5 ms, to $V_{bias}$. The bias voltage increases at 0.53 s such that $V_{bias}$ exponentially approaches 40 M V/m , while the voltage decreases at 1.78 seconds in an exponential approch towards 0 M V/m. The magnitude of the electric field sweep rate $\mid\dot{E}_{\textrm{bias}}\mid$ is shown in the Fig. \ref{fig:TempDep}(d) inset. Approximately 20 waveform cycles were averaged and a 5 ms time resolution was used to extract $\tan\delta$ at each time slice, which is shown in Fig. \ref{fig:TempDep}(d).

While the TLS density stays approximately constant as the bias is varied (according to the static measurements discussed above), Fig. \ref{fig:TempDep}(d) shows that $\tan\delta$ increases dramatically when the bias is swept, i.e. when the electric field has a significant sweep rate magnitude. Comparing Fig. \ref{fig:TempDep}(c) and \ref{fig:TempDep}(d) (with arrows shown) reveals that for both temperatures and during the fastest sweep rates, the (strong non-equilibrium) loss tangent approximately equals the linear-response equilibrium loss tangent, and it is smaller for lower sweep rates. Figure \ref{fig:TempDep}(e) shows this strong non-equilibrium $\tan\delta$ for several temperatures with the linear-response equilibrium $\tan\delta$; note that the former value is less than $5\%$ higher than the latter value. This correspondence between strong non-equilibrium  $\tan\delta$ and the equilibrium linear-response  $\tan\delta$ has been seen in multiple samples. The solid fit curve shows the expected thermal saturation for equilibrium TLS, indicating that the high sweep rate phenomena are related. We note that most of the return to the steady-state loss occurs while the bias voltage is changing at a rate $\mid\dot{E}_{\textrm{bias}}\mid> 10^{-4}$ V/m/s. For negligible sweep rates, $\mid\dot{E}_{\textrm{bias}}\mid < 10^{-4}$ V/m/s, a small amount of slow dynamics can be seen in the 33 mK data, which might possibly contain glassy relaxation phenomena, but regardless of the mechanism the 200 mK data returns to steady state equilibrium relatively fast indicating that coherence is involved in the slow 33 mK dynamics. Below we analyze the main nonequilibrium phenomenon, the loss tangent as a function of bias sweep rate, using the theory of Burin \textit{et al.} \cite{Burin2013}, which is predicted to apply (universally) to all amorphous dielectrics. 

In our system the Rabi frequency, $\Omega_R$, of a TLS can be larger than the decoherence rate for a TLS, but it is always much smaller than the resonance frequency such that multiple photon processes can be ignored. A swept electric field bias changes the TLS energy at a rate of $\hbar v=p\dot{E}_{bias}\cos(\theta)(\Delta/E_{TLS})$ \cite{Burin2013}, where $\theta$ is the angle between the field and the TLS dipole.  Below we explore the non-equilibrium loss tangent in the fast sweep regime corresponding to $\Omega_{R0}^2/v_0<<1$, where $v_0=v(\Delta=\hbar\omega, \theta=0)$ is the maximum sweep rate, $\Omega_{R0}=\Omega_R(\Delta_0=\hbar\omega, \theta=0)$ is the maximum Rabi frequency on resonance, and as discussed earlier $\Delta_0$ is the TLS tunneling rate. When $v_0$ is sufficiently small, the steady-state equilibrium loss should be recovered due to TLS decoherence processes. For an individual TLS that is swept through the ac-field frequency, the probability of an adiabatic transition of the TLS-photon field from $|g,n\rangle$ to $|e,n-1\rangle$ is $P=1-e^{-\gamma}$, where $\gamma=(\pi\Omega^2_R)/(2v)$. This creates a non-equilibrium loss tangent \cite{Burin2013} of 
\begin{equation}
\tan\left(\delta\right)=\frac{16\pi P_0}{\epsilon E^2_{ac}}\int^{\hbar\omega}_0\frac{d\Delta_0}{\Delta_0}\frac{\hbar^2v(1-e^{-\gamma})}{\sqrt{1-\left(\frac{\Delta_0}{\hbar\omega}\right)^2}}.
\label{eq:tandNE}
\end{equation}
In the fast sweep limit, $v_0\gg\Omega_{R0}^2$, the TLS pass through resonance rapidly such that they Landau-Zener tunnel to remain in their ground state $(|g,n\rangle)$ with a high probability (and excited state with a low probability). This low probability of individual excitation comes with a high rate of TLS crossings such that the loss tangent is actually higher than in steady-state equilibrium, and in agreement with theory, is approximately equal (see Fig. 1(e)) to the linear response loss tangent calculated from Fermi's golden rule.

\begin{figure}
\includegraphics[trim = 0cm 0cm 0cm 0cm, clip, scale=0.85]{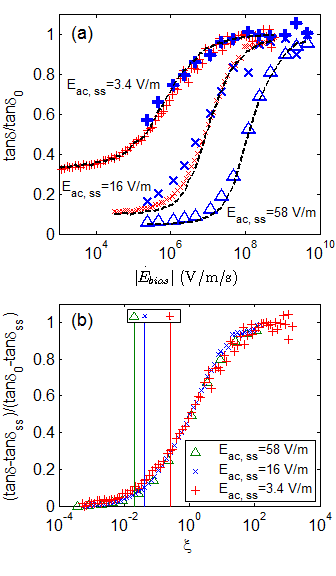}
\caption{\label{fig:SCurve} (color online) (a) Loss tangent for three different input microwave fields (+, x, $\triangle$) as a function of the bias rate $\dot{E}_{bias}$, taken with different techniques. Dashed curves are Monte-Carlo fits with dipole moment p=7.9 D. (b) Loss tangent measurements for each ac drive, normalized to their maximum and minimum values, shown as a function of the dimensionless sweep rate, $\xi$. The data collapses onto one curve.}
\end{figure}

The data points in Fig. \ref{fig:SCurve}(a) show loss tangent measurements (normalized to maximum loss tangent), as a function of the bias rate. The thick blue symbols are produced by sweeping the bias with a 8.5 ms time-constant filter (similar to shown in Fig. \ref{fig:TempDep}(d) ) of varying amplitudes, and plotting the loss tangent for the maximum bias rate (the pulse amplitude divided by the rise time). The thin red symbols are produced by using a single wave form with a (slower) 50 ms time-constant filter and plotting the instantaneous loss tangent values against the instantaneous bias rate versus time (similar to the 33 mK data of Fig. \ref{fig:TempDep}(d) versus inset quantity, but with a longer exponential tail). The data shown are taken from upward voltage steps in bias for both data types; data from downward steps in bias (not shown) are nearly indistinguishable. Measurements were performed for three different applied microwave amplitudes (+, x, $\triangle$). While the amplitude of the input microwave field is constant for any given curve, the microwave field across the capacitors is also influenced by the loss tangent; the low sweep rate (steady-state) microwave field is given as a label for each curve in Fig. \ref{fig:SCurve}(a). The microwave field for the middle curves (x) vary from $E_{ac}=4.55$ V/m at the highest bias rate to $E_{ac}=15.52$ V/m at the lowest bias rates (the steady-state regime). Eq. (\ref{eq:tandNE}) by itself is only applicable to our results at the fastest bias rates, $\xi>1$, where $\xi=(2v_0)/(\pi\Omega^2_{R0})$ is the dimensionless sweep rate, because it neglects TLS relaxation times, $T_1$ and $T_2$, which limit the loss in the steady-state regime ($\xi \ll 1$). At very slow bias rates, $\xi\ll1$, the response is adiabatic so the loss tangent approaches the steady-state loss tangent.

To model the loss for bias rates between the steady-state regime $(\xi\ll 1)$ and the strong non-equilibrium regime $(\xi \gg 1)$, we used a Monte-Carlo averaged solution of the TLS density matrix with $T_2=2T_1$. We fit the simulation to the data by varying the dipole moment (see solid curves in Fig. \ref{fig:SCurve}(a)) and use a loss tangent floor of $\tan\delta_{floor}=1.8\times 10^{-5}$ found from steady-state saturation measurements. We find excellent agreement using a single dipole moment, indicating that multiple dipole moments are not apparent in the dynamics. The fit revealed a TLS dipole moment of $p=7.9$ Debye and a value of $P_0=4.9 \times 10^{43} J^{-1}m^{-3}$ for the TLS spectral population density. The spontaneous emission limited relaxation time used in the fit is $T_{1,min}=3.0$ $\mu s$, and this can be alternatively extracted from the steady state loss tangent measurement using the extracted $p$. This moment also agrees with fits of only the fastest parts of the curves in Fig. \ref{fig:SCurve}(a) ($\xi>1$) to a numerical evaluation of Eq. (\ref{eq:tandNE}). The error from the data analysis gives an imprecision of 3\% for $p$, but the accuracy is limited to approximately 10\% by a room-temperature calibration of the ac power. The same analysis applied to a second $\mathrm{Si_3N_4}$ film type, with a different stoichiometry and linear response loss tangent $(\tan\delta_0 \simeq 10^{-4})$, yielded the same dipole moment. This $p$ is comparable to that found for TLS in an amorphous $\mathrm{Al_2O_3}$ Josephson junction tunneling barrier \cite{Martinis2005}. However, we note that unlike previous dipole measurements in tunneling barriers our technique doesn't require large statistical measurement of individual strongly coupled TLS but instead allows for ensemble TLS dipole measurement in a deposited film using a relatively fast measurement. Since this is an ensemble measurement it can be compared a previous measurements of a bulk (but not deposited) $\mathrm{SiO_2}$ insulator \cite{Golding1979}. 

In Fig. \ref{fig:SCurve}(b) we plotted one curve from each of the three ac drives in Fig. \ref{fig:SCurve}(a), scaling the bias rate on the x-axis by the parameter $\xi$ and scaling the y-axis by subtracting the constant bias steady-state loss tangent value $(\tan\delta_{ss})$ for each of the three sets of microwave amplitudes to account for the different steady-state losses.  Here the dimensionless sweep rate $\xi$ is calculated for the same extracted dipole moment, $p=7.9$ D. The location of the step in loss tangent on the x-axis of Fig. \ref{fig:SCurve}(b) is expected from the theory. However, the collapse of the data to one curve shows that the scaled loss tangent (y-axis) quantity allows a check of the dipole moment without a Monte-Carlo simulation. In Fig. \ref{fig:SCurve}(b) the vertical lines represent critical bias rate (coded by symbols and color), $\xi_c=1/(\Omega_{R0}\sqrt{T_{1,min}T_2})$, above which the TLS dynamics is described by Landau-Zener theory and is not limited by TLS relaxation times. The model also predicts that when the bias rate is slowed such that the loss tangent is reduced to below $\tan\delta_0/2$, TLSs are continuously population inverted as they pass through the ac field frequency. Therefore data above $\xi_c$ and with a loss tangent below $\tan\delta_0/2$ represent a regime where TLS passing through resonance are inverted.

In conclusion, we have used a swept electric field to continuously spectrally-tune the broad distribution of TLS bath states in an insulating film within a microwave circuit. The swept bias field is applied in a complimentary way to the microwave field, and by utilizing the Landau-Zener effect a continuous population control near resonance is achieved, as monitored by the microwave loss tangent. Under application of a fast-swept bias field and a moderate microwave drive excitation, the population is primarily left in the ground state and the measured non-equilibrium loss tangent becomes approximately equal to the equilibrium (i.e. linear response) loss tangent. As the sweep rate is lowered or the ac power is increased, coherent population inversion occurs for TLS states in the bath passing through resonance. We find good agreement with a theory based on the Landau-Zener effect and apply this theory to extract the TLS dipole moment and density of states which characterizes the TLS bath dynamically. This technique presents the first TLS dipole moment measurement in a deposited film. The TLS bath can be manipulated for long periods of time, allowing one to create new environments on chip through population control. Although the TLS population is only controlled for TLS passing through an exciting field, a fast bias rate may also invert the TLS population over a useful spectral range of frequencies. Furthermore, using the TLS bath in this way could possibly allow for lasing from a disordered set of inverted states \cite{Burin2014SST}.

\bibliography{References2}

\end{document}